\documentstyle[12pt,psfig]{article}
\def\beq{\begin{equation}}
\def\eeq{\end{equation}}
\def\bea{\begin{eqnarray}}
\def\eea{\end{eqnarray}}
\def\bq{\begin{quote}}
\def\eq{\end{quote}}

\def\gappeq{\mathrel{\rlap
{\raise.5ex\hbox{$>$}}
{\lower.5ex\hbox{$\sim$}}}}
\def\lappeq{\mathrel{\rlap{\raise.5ex\hbox{$<$}}
{\lower.5ex\hbox{$\sim$}}}}
\def\simlt{\stackrel{<}{{}_\sim}}
\def\simgt{\stackrel{>}{{}_\sim}}

\begin{document}
\pagestyle{empty}
\begin{flushright}
{CERN-TH/99-133\\
IFT-98/20\\
hep-ph/9905436}
\end{flushright}
\vspace*{5mm}
\begin{center}
{\bf
IMPLICATIONS OF THE PRECISION DATA FOR\\
VERY LIGHT HIGGS BOSON SCENARIOS IN 2HDM(II)}
\\
\vspace*{1cm} 
Piotr H. Chankowski$^{a,b)}$, Maria Krawczyk$^{b)}$ and Jan \.Zochowski$^{b)}$
\\
\vspace*{1.7cm}  
{\bf ABSTRACT} \\
\end{center}
\vspace*{5mm}
\noindent
We present an up-to-date analysis of the constraints imposed bythe precision 
data on the ($CP-$ conserving) Two-Higgs-Doublet Model of type II, with 
emphasis on the possible existence of very light neutral (pseudo)scalar 
Higgs boson with mass below 20--30 GeV. We show that even in the presence 
of such light particles, the 2HDM(II) can describe 
the electroweak data with precision comparable to that given by the SM.
Particularly interesting lower limits on the mass of the lighter neutral
$CP-$even scalar $h^0$ are obtained in the scenario with a
light $CP-$odd Higgs boson $A^0$ and large $\tan\beta$. 
\vspace*{1.0cm} 
\noindent

\rule[.1in]{16.5cm}{.002in}
\noindent
$^{a)}$ Theory Division, CERN, Geneva, Switzerland \\
$^{b)}$ Institute of Theoretical Physics, Warsaw University, Poland \\
\vspace*{0.2cm}

\begin{flushleft} 
CERN-TH/99-133\\
IFT-98/20\\
May 1999
\end{flushleft}
\vfill\eject
\newpage

\setcounter{page}{1}
\pagestyle{plain}

\section{Introduction}
The  Standard Model (SM) of electroweak interactions is in very good 
agreement with the electroweak precision data collected at LEP and SLAC 
experiments \cite{LEPEWWG}. This fact strongly supports the idea of the 
spontaneous breaking of the underlying $SU_L(2)\times U_Y(1)$ gauge symmetry.
Yet, the actual mechanism of symmetry breaking still remains unexplored. 
The Higgs boson predicted by the minimal model of electroweak symmetry 
breaking, the SM, has not been found up to now. Only the
lower limit on its mass of $\sim90$ GeV is set by the unsuccessful search at 
LEP. While the extreme simplicity of the Higgs sector of the SM is 
theoretically appealing, there exist many of its extensions that lead to 
different phenomenology (more physical Higgs particles) and which also 
should be tested (or constrained) experimentally.

The simplest such extension is the well-known Two-Higgs-Doublet model 
(2HDM). It exists in several distinct versions, of which we want to consider 
in this article the one that is called Model II in its $CP-$conserving version
(we briefly recall its structure in the next section).\footnote{The Higgs 
sector of the Minimal Supersymmetric Standard Model has precisely the 
structure of Model II, but with additional constraints imposed on the 
quartic couplings. However, in the following we will consider the general 
Model II in a regime in which it cannot be regarded as a low-energy 
approximation of the MSSM with heavy sparticles.} 
One interesting question that arises in the context of such an extension 
of the SM is what the available experimental limits are on the masses of the 
Higgs bosons predicted in such a model. In Section~3 we will recall the 
arguments \cite{MK} that, in the framework of the considered version of 
2HDM, the direct searches do not exclude the existence of very light 
neutral scalar or pseudoscalar Higgs particles. In Section 4 we show that 
the existence of such light Higgs bosons is notexcluded by the electroweak 
precision data either. An analysis of the impact of the precision data 
on the 2HDM(II) was already performed in the past \cite{COHOMO,GRANT} 
(for the formalism and early investigations, see also \cite{HO}) but 
concentrated mainly on the possibility of improving the prediction for 
$R_b\equiv\Gamma(Z^0\rightarrow\bar bb)/\Gamma(Z^0\rightarrow hadrons)$ 
(which at that time seemed to be required by the data) and on improving
the 2HDM(II) global fit to the data with respect to the fit given by the 
SM. Since then, the  experimental situation has evolved  significantly. In
particular the  measurement of $R_b$ no longer shows any statistically
significant deviation from the value predicted by the SM \cite{LEPEWWG}. 
Also there are
changes both in the experimental measurement and theoretical computation
of the $b\rightarrow s\gamma$ decay rate, which was crucial in the analysis 
performed in ref. \cite{GRANT}. More recently a partial analysis of the 
constraints imposed on the 2HDM(II) by various measurements was also 
attempted in \cite{HALO}. Here we present an up-to-date analysis of the 
constraints that the precision data impose on the 2HDM(II), with emphasis on 
the possible existence of light neutral (pseudo)scalar Higgs boson. We show
that even in the presence of such light particles, the 2HDM(II) can describe 
the electroweak data with the precision comparable to that given by the SM. 
In this case some interesting global limits on the model can be obtained.
Finally in Section 5 we summarize our results and briefly comment on the other 
ways the existence of light Higgs bosons are or can be constrained by other 
experimental data.
\vskip 0.3cm

\section{Two-Higgs-Doublet extension of the SM - Model II}

The multidoublet extensions of the SM are distinguished
by their virtue of not introducing corrections to the $\rho$ parameter
at tree level. The minimal extension of the SM consists of two doublets. 
The requirement of the absence at tree level of the flavour-changing 
neutral currents puts restrictions on how the two scalar doublets
of the general 2HDM can couple to fermions. In Model II, 
one Higgs doublet (denoted by $\Phi_1$) couples only to leptons and down-type 
quarks, whereas the other doublet ($\Phi_2$) couples only to up-type quarks. 
After spontaneous symmetry breaking, both doublets acquire vacuum expectation 
values $v_1$ and $v_2$, respectively with $v\equiv\sqrt{v_1^2 + v_2^2}$ 
fixed by $M_W$, and 
\begin{eqnarray}
{v_2\over v_1} \equiv\tan\beta.
\end{eqnarray}
With two complex Higgs doublets, the 2HDM predicts the existence of five 
physical scalars: neutral $h^0$, $H^0$ and $A^0$ and charged $H^\pm$.
In the more restrictive scenario (which we are going to discuss),
with $CP$ symmetry conserved by the Higgs potential,
$h^0$ and  $H^0$ are $CP-$even mixtures of the neutral components of the
doublets (the mixing being parametrized by the angle $\alpha$) whereas
$A^0$ is $CP-$odd. Thus, in the $CP-$conserving version, the Higgs sector
is parametrized by four masses $M_h$, $M_H$ (by definition $M_h\leq M_H$),
$M_A$ and $M_{H^+}$ and three dimensionless parameters $\tan\beta$, 
$\alpha$ and $\lambda_5$ (for definitions see for instance, \cite{HHG}). 
The first 
two dimensionless parameters are very important for the phenomenology of the 
Higgs sector as they determine the couplings of the physical Higgs bosons
to fermions and gauge bosons: the couplings of the scalars to the 
down- and up-type quarks are given by the SM couplings multiplied by the
factors (see for instance, \cite{HHG,HABER})
\begin{eqnarray}
h^0b\bar b:&& -{\sin\alpha\over\cos\beta}
=\sin(\beta-\alpha)-\tan\beta\cos(\beta-\alpha)\nonumber\\
h^0t\bar t:&&  {\cos\alpha\over\sin\beta}
=\sin(\beta-\alpha)+\cot\beta\cos(\beta-\alpha)\nonumber\\
H^0b\bar b:&&  {\cos\alpha\over\cos\beta}
=\cos(\beta-\alpha)+\tan\beta\sin(\beta-\alpha)\nonumber\\
H^0t\bar t:&&  {\sin\alpha\over\sin\beta}
=\cos(\beta-\alpha)-\cot\beta\sin(\beta-\alpha).
\label{eqn:higffcpl}
\end{eqnarray}
The Feynman rules for the $CP-$odd 
scalar couplings to fermions are given by the SM rules for $h^0_{SM}$ 
times the factors:
\begin{eqnarray}
A^0b\bar b:~ -i\gamma^5\tan\beta, ~~~
A^0t\bar t:~ -i\gamma^5\cot\beta.\label{eqn:affcpl}
\end{eqnarray}
Important for the direct Higgs boson search at LEP, couplings $Z^0Z^0h^0$
and $Z^0Z^0H^0$ are given by the corresponding Standard Model 
coupling $Z^0Z^0h^0_{SM}$ modified by the factors:
\begin{eqnarray}
Z^0Z^0h^0:~~ \sin(\beta-\alpha), ~~~~
Z^0Z^0H^0:~~ \cos(\beta-\alpha),\label{eqn:zhcouplings}
\end{eqnarray}
whereas the couplings of $Z^0$ to $A^0h^0$ and $A^0H^0$ pairs are 
instead proportional to
\begin{eqnarray}
Z^0A^0h^0:~~ \cos(\beta-\alpha), ~~~~
Z^0A^0H^0:~~ \sin(\beta-\alpha).
\label{eqn:ahcouplings}
\end{eqnarray}
It follows that, in the limit $\sin(\beta-\alpha)=1$, the lighter $CP-$even
neutral Higgs boson $h^0$ has precisely the couplings of the Standard Model 
Higgs and becomes indistinguishable from it. Finally, couplings of the 
charged Higgs scalar to fermions, e.g. $\bar btH^-$ vertex, are given by
expressions like:
\begin{eqnarray}
{g\over2\sqrt2M_W}\left[m_t\cot\beta(1+\gamma_5)
+ m_b\tan\beta(1-\gamma_5)\right].\nonumber
\end{eqnarray}
It is clear that for $\tan\beta$ close to zero, scalar couplings to $t\bar t$ 
pairs are strongly enhanced with respect to to the SM case, whereas for
$\tan\beta$ large ($\simgt10$) the couplings to $b\bar b$ are enhanced. 
Requirement of perturbativity of both types of couplings restricts, therefore, 
$\tan\beta$ values to the range \cite{LIMITTG}:
\begin{eqnarray}
0.3\simlt\tan\beta\simlt130.
\label{eqn:tbrange}
\end{eqnarray}
Outside this range perturbativity is lost and no firm prediction 
can be obtained from the model.

\begin{figure}
\psfig{figure=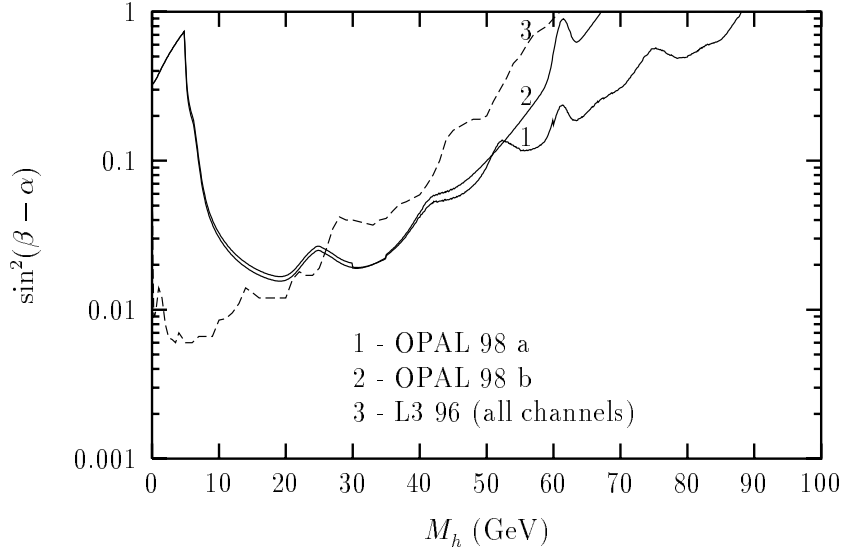,width=12.0cm,height=8.0cm} \vspace{1.0truecm}
\caption{$95\%$ C.L. limits on $\sin^2(\beta-\alpha)$ from the Higgs
boson search at LEP as a function of $M_h$. Lines 1 and 2 are the  
OPAL results obtained assuming the Higgs boson decay branching ratios
as in the SM and 100\% into hadrons, respectively 
\protect\cite{OPALSIN}. Line 3 is the 
result of L3 \protect\cite{SINUSL3}.}
\label{fig:sinlim}
\end{figure}

\vskip 0.3cm

\section{Constraints on the 2HDM(II) from direct Higgs boson search at LEP}

\begin{figure}
\psfig{figure=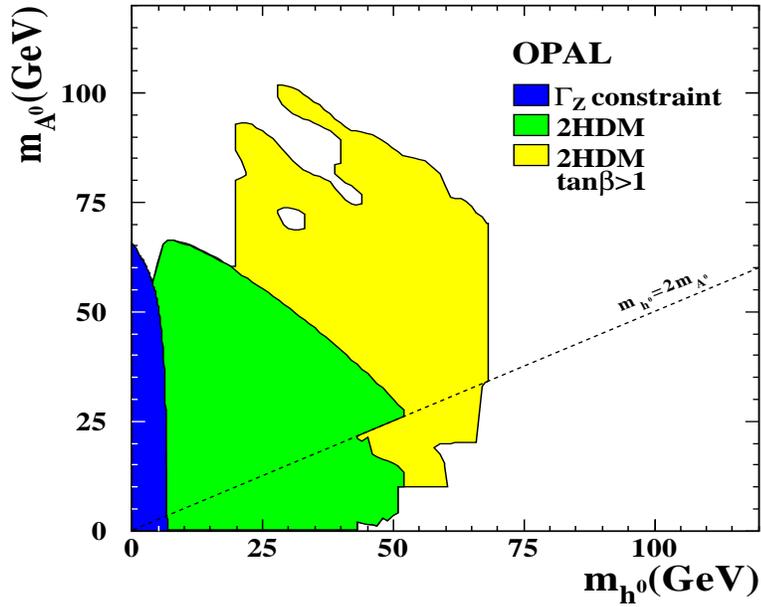,width=10.0cm,height=8.0cm} \vspace{1.0truecm}
\caption{$95\%$ C.L. limits on the $(M_h, ~M_A)$ plane from the Bjorken 
process and the associated Higgs boson production as given by the OPAL 
collaboration \protect\cite{ALEPH}.}
\label{fig:mhmalim}
\end{figure}

Here we briefly review the constraints on the 2HDM(II) imposed by
direct Higgs boson search at LEP \cite{MK}, which will be taken into account 
in performing the fits to the electroweak data in Sec.~4.
At the LEP collider, the Higgs boson search is based mainly on the 
following processes:
\begin{enumerate}
\item the Bjorken process $e^+e^-\rightarrow Z^{0\star} h^0(H^0)$ 
\item the associated Higgs boson pair production 
$e^+e^-\rightarrow A^0h^0(H^0)$ 
\item the Yukawa process $e^+e^-\rightarrow b\bar b\rightarrow b\bar bA^0(h^0)$
\item $e^+e^-\rightarrow H^+H^-$.
\end{enumerate}
Because of the structure of the $Z^0Z^0h^0$ and $Z^0A^0h^0$ couplings 
(\ref{eqn:zhcouplings}),(\ref{eqn:ahcouplings}) the 
processes $e^+e^-\rightarrow Z^{0\star} h^0$ and $e^+e^-\rightarrow A^0h^0$
are complementary to each other, provided they are simultaneously 
kinematically allowed.
In the SM, or in the 2HDM(II) in the limit $\sin(\beta-\alpha)=1$, the 
non-observation of the Bjorken process at LEP sets a lower limit on 
$M_{h_{SM}}$ of $\sim90$ GeV (at $95\%$ C.L.) \cite{LEPHiggs}. In the 
general case, the same data put an upper limit on the factor 
$\sin^2(\beta-\alpha)$ as a function of the lighter Higgs boson mass $M_h$ 
(see Fig.~\ref{fig:sinlim}). The combined analysis of the complementary 
channels 1 and 2, performed for instance, 
by the OPAL collaboration \cite{OPALSIN}, leads to the constraints on the 
$(M_h, ~M_A)$ plane shown in Fig.~\ref{fig:mhmalim}. It follows that 
the direct experimental limits on $M_h$ and $M_A$ are rather weak: the data
allow for very light $h^0$ ($A^0$) provided $A^0$ ($h^0$) is heavier than 
$\sim65$ (50) GeV even if only $\tan\beta>1$ is allowed. In particular,
there exist no absolute bound on $M_h$ from LEP data, provided one 
respects the bound on $\sin^2(\beta-\alpha)$ shown in Fig.~\ref{fig:sinlim}. 

\begin{figure}
\psfig{figure=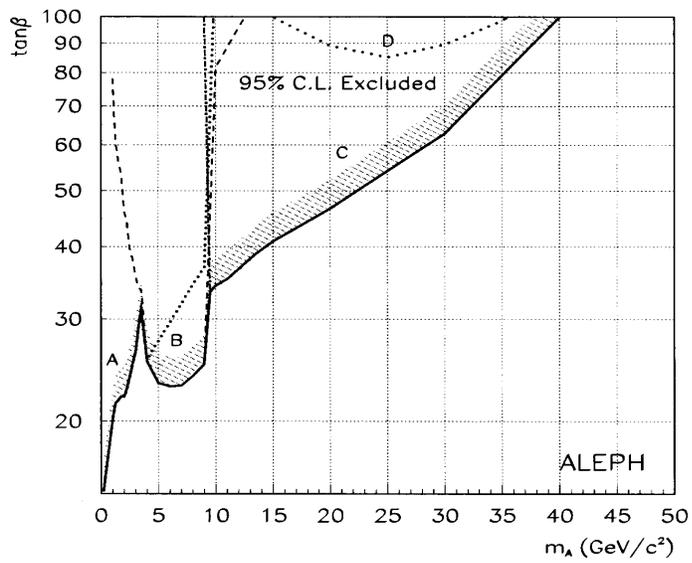,width=10.0cm,height=8.0cm} \vspace{1.0truecm}
\caption{$95\%$ C.L. limits on the $(\tan\beta, ~M_A)$ plane from the 
Yukawa process as given by the ALEPH collaboration \protect\cite{ALEPH}.}
\label{fig:tbmalim}
\end{figure}

For large values of $\tan\beta$, independent constraints on $h^0$ and $A^0$ 
follow from the non-observation of the Yukawa process. The available limits 
on the $(M_A, \tan\beta)$ plane are shown in Fig.~\ref{fig:tbmalim} 
\cite{ALEPH}. Similar limits on both $(M_h, \tan\beta)$ and $(M_A,\tan\beta)$ 
planes have been reported only recently by the DELPHI Collaboration 
\cite{ZALEWSKI}. 

It should also be mentioned that some, rather weak 
(and dependent on the assumptions made about the Higgs boson decay
branching fractions) limits on the very light $h^0$ and 
$A^0$ can be derived from the so-called Wilczek processes \cite{WILCZEK},
i.e. from $\Upsilon$ and $J/\psi$ decays into $h^0(A^0)$ and the photon 
\cite{UPSILON}. 
We do not take these limits into account in performing the fit to the 
electroweak data, because the results we will show do not change as $M_h$ 
varies over the range 0--30 GeV. It should also be clear that 
whenever $\sin^2(\beta-\alpha)\approx0$ the heavier $CP-$even scalar, 
$H^0$, is constrained by the Bjorken process so that $M_H\simgt90$ GeV.
Finally, the non-observation of the charged Higgs boson production at LEP
sets the bound $M_{H^+}>72$ GeV \cite{HPLUSLIM}, which is, however, much 
less restrictive than the indirect limit derived in the 2HDM(II) from the 
$b\rightarrow s\gamma$ process and, for $\tan\beta<1$, from $R_b$.

{}From the above, it follows that within the 2HDM(II) there still exist two 
scenarios with either very light scalar $h^0$ or very light pseudoscalar 
$A^0$, which are not excluded by the available data (for other 
constraints not discussed here, see \cite{MKAB,MKJK} and Section 5). 

In the next section we will consider
how the electroweak precision data constrain these two scenarios.

\vskip 0.3cm 

\section{Global fit to the precision data}

Since the advent of  LEP  and SLAC experiments, precision  electroweak
data have been playing an increasingly important r\^ole  in  constraining the
mass of  the top quark in the  Standard Model (SM). Nowadays, with the
top quark mass   directly measured  at Fermilab \cite{TOPMASS},   they
significantly constrain the last unknown parameter of the SM, the mass
of  the  Higgs  boson  \cite{ELFOLI}. They  are   also very useful in
constraining  possible   forms of new   physics  such as supersymmetry
\cite{CHPO}  or  technicolour \cite{ALT}. It   is, therefore, natural 
to ask to what extent the precision electroweak data constrain the Two  
Higgs Doublet Model of the type II. In  this context, a question of
particular  importance is whether they  are  still compatible with the
existence of a very light neutral Higgs particle (scalar or pseudoscalar 
one) which, as we discussed in the preceding Section, is not yet excluded 
by direct searches. In this  Section we discuss  under
what  conditions  the  existence of    light $h^0$   or  $A^0$ can  be
compatible   with indirect   constraints  imposed  by  precision data.

A strictly statistical  approach to  constraining  the  2HDM(II)  indirectly
would consist of finding  the global minimum  of the $\chi^2$ fit and,
in  the next step,  excluding  all points in   the parameter space for
which $\Delta\chi^2$  is greater than 3.84  (exclusion at  95\% C.L.).
Comparison of such an analysis with the similar one carried for the SM
would reveal that, per degree of freedom (d.o.f.), the  fit in 2HDM(II) is
much worse than  in  the SM.  This   follows from  the fact that   the
description  of the electroweak data by  the  latter is nearly perfect
\cite{LEPEWWG}. Hence, a 2HDM(II) predicting  individual observables as
accurately as the SM would  have much worse $\chi^2/$d.o.f., because of
larger number of free parameters in  its Higgs potential.  
In our investigation we
do  not follow such an approach.  Rather, we take   the SM best global
$\chi^2$ value as a reference point and concentrate on the qualitative
discussion of which regions of the 2HDM(II) parameter space can give an
equally  good global    $\chi^2$  value. In  order    not  to be   too
restrictive, when constraining  the parameter space  we use  the rough
criterion that the $\chi^2$ in the 2HDM(II) should not be greater than the
SM best $\chi^2$ value  plus 4. Our  emphasis is,  however, on  the fact
that such a criterion  (in fact any reasonable  one) does allow  for a
very light neutral scalar or pseudoscalar. The data we take into account 
include the precision
electroweak data reported at the Moriond '98 conference \cite{LEPEWWG},
which are  dominated by  those  from  LEP~1. For  future  reference we
record that the best SM  fit to the   electroweak data we have  chosen
gives us $\chi^2\approx15.5$. Therefore,  all bounds on the  masses of
the  2HDM(II) Higgs bosons  we will present  are derived  by requiring the
$\chi^2$ in that model to be less than 19.5. 

In discussing the values taken by the $\chi^2$ for various Higgs boson
mass configurations (to explain qualitatively the origins of the bounds 
we show) we will always start with a discussion of the contribution
to the $\Delta\rho$ parameter, defined as 
\begin{eqnarray}
\Delta\rho={\Pi_{WW}(0)\over M_W^2} - {\Pi_{ZZ}(0)\over M_Z^2}
          - 2{s_W\over c_W}{\Pi_{Z\gamma}(0)\over M_Z^2},
\end{eqnarray}
where $s_W ~(c_W)$ is the sine (cosine) of the Weinberg angle.
It largely determines the bulk of the    
predicted values of the electroweak  observables such as $M_W$,
$\sin^2\theta^{eff}$ (measured   through   various asymmetries  of the
final  fermions and/or their  polarizations  \cite{LEPEWWG}),  etc., 
and is therefore the main factor shaping the $\chi^2$ curves, at least
for not too small or too large values of $\tan\beta$, i.e. when the 
couplings of the Higgs bosons to the $b\bar b$ pair are not enhanced.

In addition, for low and high values of $\tan\beta$, a particularly important 
r\^ole is played in the fit by the quantity $R_b$. The current experimental 
result is $R_b=0.21656$, with error
$\Delta R_b=0.00074$ (which is 0.9 standard deviation
above the SM prediction). Its importance follows from the fact that, in 
the 2HDM(II), the contribution of the Higgs bosons can easily change the 
prediction for $R_b$ with respect to the SM, spoiling the $\chi^2$ fit to the 
data. This contribution has been  studied in detail in ref. \cite{DENNER}. 
(Handy formulae  for $\delta R_b$ can also be found in the Appendix of 
ref. \cite{CHPORb}.) The contribution of $H^+$ to $\delta
R_b$ contains parts that are proportional to $(m_t/M_Z)^2\cot^2\beta$ and
$(m_b/M_Z)^2\tan^2\beta$ and, consequently, can be sizeable for either
very small  or very large values  of $\tan\beta$.  The neutral scalars
become relevant to $R_b$ only in the latter limit 
(see. Eqs.~(\ref{eqn:higffcpl}) and (\ref{eqn:affcpl})).  For a qualitative
understanding of the  results it  is sufficient  to remember that  the
contribution of $H^+$ is  always negative, whereas the contribution of
the neutral  Higgs  bosons can be positive  provided  $A^0$ is  not too
heavy, say $M_A\simlt100$ GeV, and the splitting  between its mass and the
mass of the $CP-$even  scalar $h^0$ or $H^0$ (the one  which, for a given 
angle $\alpha$,
couples more strongly to the $b\bar b$ pair) is not  too large. 
For example, in the configuration (typical for the Minimal SUSY Standard
Model with large $\tan\beta$) $M_h\sim M_A\simlt 70$ GeV (and 
$\sin^2\alpha\approx1$) the contribution of $A^0$  and $h^0$  can easily 
overcompensate the  negative
contribution of $H^+$ with mass $M_{H^+}>100$ GeV. 
It is the interplay of the contribution to the $\Delta\rho$ parameter
and the contribution to $\delta R_b$ that is responsible for interesting
bounds on 2HDM(II) with light (pseudo)scalar particle, which can be derived
on the basis of the $\chi^2$ fit.

Another very important  constraint on new physics is the  measured   
value of the branching ratio $BR(B\rightarrow X_s\gamma)$ \cite{CLEOnew}.  
In the context of the 2HDM(II) this measurement can be converted into a lower 
bound on the mass of the charged Higgs boson \cite{BUMIMUPO}. 
Recently a big effort was made by various groups to improve the accuracy 
of the theoretical prediction for this ratio \cite{ADYA,CIDEGAGI,NEU}. Our 
lower bound on $M_{H^+}$, based on $b\rightarrow s\gamma$ for $m_t=174$ GeV,
is shown as a 
function of $\tan\beta$ in Fig.~\ref{fig:lhglobal}a by the solid line.
\footnote{We compute $BR(b\rightarrow s\gamma)$ with NLO accuracy, following 
the approach of ref.~\cite{MIK} supplemented with electromagnetic as well 
as $1/m_b^2$ and $1/m_c^2$ corrections setting the parameter $\delta=0.9$
\cite{NEU}. The theoretical 
uncertainty is taken into account by computing the rate for $\mu_b=2.4$ and 
9.6 GeV and then shifting its larger (smaller) value upward (downward)
by the errors, added in quadrature, related to the uncertainties in 
$\alpha_s$, $m_b$, $m_c/m_b$, $|V_{tb}V^\star_{ts}/V_{cb}|^2$, and 
higher-order electroweak corrections; we do not take into account the 
variation of 
the scale $\mu_W$. If the resulting band of theoretical predictions
for $BR(b\rightarrow s\gamma)$ has an overlap with the CLEO 95\% C.L. band,
the point is allowed.}
It is (for large $\tan\beta$) higher by some 35 GeV than the recent estimate 
\cite{GRBO}, which gives $M_{H^\pm}\simgt165$ GeV. In view of the well-known
exquisite sensitivity of the limit on $M_{H^+}$ to the details of the analysis,
this should be considered as a satisfactory agreement. We will, however, try 
to  keep  open the possibility  that 
the charged Higgs boson can be as light as $\sim165$ GeV. 
  
\begin{figure}
\psfig{figure=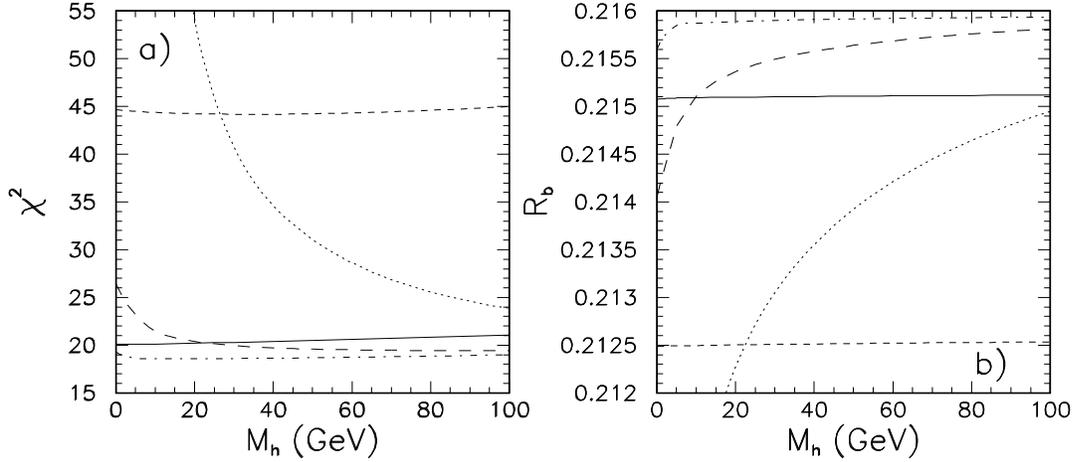,width=15.0cm,height=7.0cm} \vspace{1.0truecm}
\caption{$\chi^2$ and $R_b$ as a function of $M_h$ in the case of triple
degeneracy with $M_D\equiv M_H=M_A=M_{H^+}=250$ GeV and
$\sin^2(\beta-\alpha)=0$. Short-dashed, 
solid, dot-dashed, long-dashed and dotted lines correspond to $\tan\beta=$0.5, 
1, 5, 20 and 50, respectively. The top mass is fixed to $m_t=174$ GeV.}
\label{fig:tripledeg}
\end{figure}

\subsection{Light $h^0$}

We discuss the light $h^0$ scenario first.  For the sake of clarity it
is convenient to distinguish two $h^0$ mass ranges:
{\it i)} $M_h\simlt30$  GeV and {\it ii)}  $M_h > 30$ GeV.  
This division follows from 
the upper bounds imposed by the direct LEP search on the allowed value  
of $\sin^2(\beta-\alpha)$  \cite{SINUSL3} (Fig.~\ref{fig:sinlim}):  in  
the  case {\it i)} $\sin^2(\beta-\alpha)<0.01$--0.02 (and for practical 
purposes can be set to zero); in the case {\it ii)} the upper bound on 
$\sin^2(\beta-\alpha)$ changes roughly linearly (on the logarithmic scale) 
{}from $\sim0.02$ for  $M_h\approx30$ GeV up to
1  for $M_h\approx90$  GeV (see Fig.~\ref{fig:sinlim}).  In this paper
we will  be  mainly interested in case {\it i)} and will essentially
not explore the case {\it ii)}, which requires a more involved analysis. 
It will also  prove helpful  to  consider separately
two   ranges   of the  parameter   $\tan\beta\equiv v_2/v_1$, namely
$0.5\simlt\tan\beta\simlt10$--20 (small and intermediate) and
$20\simlt\tan\beta\simlt50$ (large).  Distinct
properties of these regions follow from the different sensitivity of
the important  observable $R_b$ to the masses  of the Higgs bosons for
small and intermediate values of $\tan\beta$, and for large ones. 

\begin{figure}
\psfig{figure=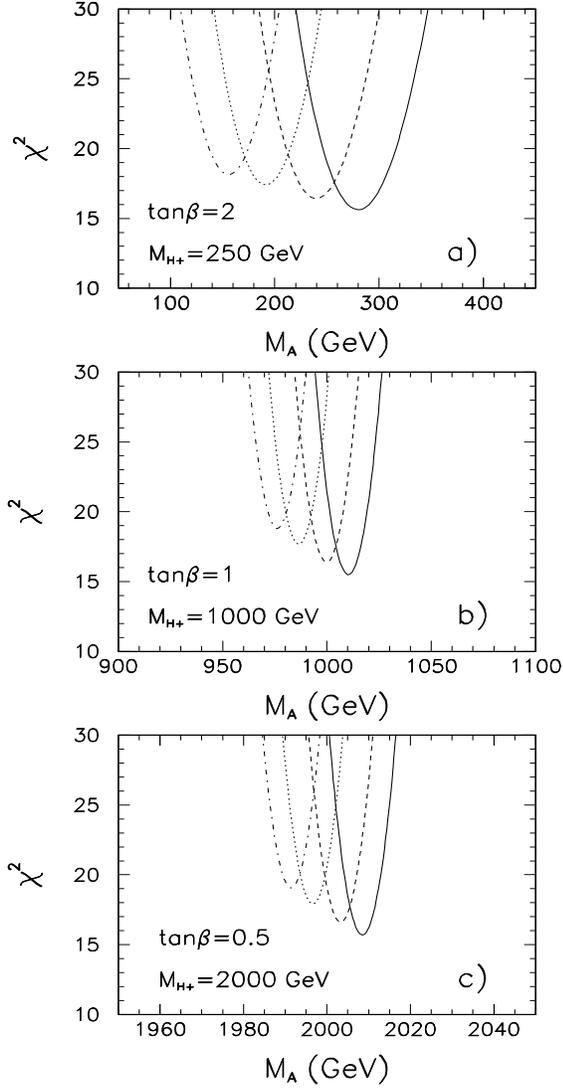,width=8.0cm,height=15.0cm} \vspace{1.0truecm}
\caption{$\chi^2$ for $M_h=20$ GeV and $\sin^2(\beta-\alpha)=0$ as a 
function of $M_A$ for different low and intermediate values of $\tan\beta$ 
and different $H^+$ masses. Solid, dashed, dotted and dot-dashed lines 
correspond to $M_H=$90, 200, 500 and 1000 GeV, respectively; $m_t=174$ GeV.}
\label{fig:lhsmtb}
\end{figure}

In order to understand the main features of the $\chi^2$
fit to the electroweak data qualitatively, 
it is instructive to begin the discussion  of the light $h^0$ case
with the somewhat peculiar limit in which  the remaining Higgs bosons,
$H^\pm$, $A^0$ and  $H^0$, are  exactly degenerate  and have a  common
mass  $M_D$. In Fig.~\ref{fig:tripledeg}a  we show the  value of 
$\chi^2$ obtained in the 2HDM(II) as a  function of $M_h$ for different  
values of  $\tan\beta$ and $M_D=250$ GeV. Corresponding
values of $R_b$  are shown in Fig.~\ref{fig:tripledeg}b. In both
cases we set $\sin^2(\beta-\alpha)=0$ and  we  keep $m_t$ fixed 
at 174 GeV.  The pattern observed in Fig.~\ref{fig:tripledeg}a can 
be easily understood by checking the contributions    to  the  parameter
$\Delta\rho$, which can be represented, in the 2HDM \cite{HHG},
in the convenient form:
\begin{eqnarray}
\Delta\rho = {\alpha\over4\pi s^2_WM^2_W}A(M_A,M_{H^+}) 
+ \cos^2({\beta-\alpha}) \Delta_c+ \sin^2({\beta- \alpha}) 
\Delta_s, \label{eqn:drho}
\end{eqnarray}
where 
\begin{eqnarray}
\Delta_c= {\alpha\over4\pi s^2_WM^2_W}\left[A(M_{H^+},M_h) - A(M_A,M_h)\right]
        + \Delta\rho_{SM}(M_H)\label{eqn:drhoc}
\end{eqnarray}
and $\Delta_s$ is obtained  by the exchange  $M_h\leftrightarrow M_H$.
The function
\begin{eqnarray}
A(x,y)=A(y,x)\equiv {1\over8}x^2+{1\over8}y^2 
- {1\over4}{x^2y^2\over x^2-y^2}\log{x^2\over y^2}
\end{eqnarray}
is positive and large if $x\gg y$ and vanishes for $x=y$. Finally,
\begin{eqnarray}
\Delta\rho_{SM}(M)&=&{\alpha\over4\pi s^2_WM^2_W}
\left[A(M,M_W) - A(M,M_Z)\right]\nonumber\\
&+&{\alpha\over4\pi s^2_W}\left[{M^2\over M^2-M^2_W}\log{M^2\over M^2_W}-
{1\over c_W^2}{M^2\over M^2-M^2_Z}\log{M^2\over M^2_Z}\right]
\label{eqn:drhosm}
\end{eqnarray}
is the Standard Model Higgs boson contribution to $\Delta\rho$. A good 
quality of the fit is obtained for 
\begin{eqnarray}
\Delta\rho_{NEW}=\Delta\rho - \Delta\rho_{SM}(M_{h^0_{SM}})\approx0,
\label{eqn:drhonew}
\end{eqnarray}
where $M_{h^0_{SM}}$ is a reference SM Higgs boson mass $\approx100$ GeV.

Since  in the case of triple degeneracy shown in Fig.~\ref{fig:tripledeg}
$\Delta\rho$ is independent of $\tan\beta$,
$\sin^2(\beta-\alpha)$ and $M_h$ values, for fixed $M_D$ the $\chi^2$
curves reflect mainly \footnote{For a qualitative  explanation of the
shapes  of the $\chi^2$ curves  the other  ``oblique'' parameters, $S$
and $U$ \cite{STU}, play  only a secondary r\^ole.}  the  dependence  
of $R_b$ on $M_h$ for different values of $\tan\beta$. For low and 
moderate values of $\tan\beta$ the predicted value of $R_b$   
is $M_h$-independent, whereas for large $\tan\beta$
($\simgt20$), for which the $h^0b\bar b$ coupling is enhanced (see 
Eq.~(\ref{eqn:higffcpl})), the  sensitivity  of  $R_b$ to  $M_h$
becomes  crucial. In  addition, for light  $H^+$ (i.e. small $M_D$,
$\sim200$ GeV) its  negative
contribution to $R_b$ spoils  the $\chi^2$ fit  for very small or very
large values of $\tan\beta$ (but this effect is $M_h$-independent). On
top of that comes the global sensitivity of the fit to
$M_D$, which enters  through $\Delta\rho_{SM}(M_D)$. This dependence is
exactly the same as the dependence of the SM fit  (for fixed top quark
mass) on  the value of  the SM Higgs boson mass.   It follows that the
best fit  (in the case of triple  degeneracy) is obtained for $M_D$ as
low as possible and intermediate values of $\tan\beta$. Note that only
the case of $M_D=250$ GeV and $\tan\beta=5$ (and marginally, for heavier
$h^0$, also for $\tan\beta=20$) shown in Fig.~\ref{fig:tripledeg}
yields an acceptable value of  $\chi^2$.  Allowing for  $M_D=165$ GeV gives
(for intermediate  values of $\tan\beta$) $\chi^2\approx17.5$.  In the
region {\it i)}  taking $\sin^2(\beta-\alpha)=0.01$  improves $\chi^2$
only marginally  (it decreases by  $\sim0.2$ for $M_D=250$ GeV).
In  the region {\it ii)} taking the largest value of $\sin^2(\beta-\alpha)$
allowed by the LEP data, e.g. $\sin^2(\beta-\alpha)=0.2$  for  $M_h=50$, 
GeV decreases  for $M_D=250$ GeV the $\chi^2$
value to  $\sim18.1$ as a result of a ``redistribution'' of contributions
in the last two terms in Eq.~(\ref{eqn:drho}).

Essential  improvement of the   fit   is,  however, obtained  by   the
departure from  the strict limit  of triple degeneracy. For very small
and intermediate values of $\tan\beta$,  this is illustrated in
Fig.~\ref{fig:lhsmtb} where  for $M_h=20$ GeV (which is representative of 
$0<M_h\simlt30$ GeV), $\sin^2(\beta-\alpha)=0$, and $m_t$ fixed  at 174 GeV,
we  show $\chi^2$ as a function of $M_A$ 
for  different combinations of $H^+$ and $H^0$ masses. 

\begin{figure}
\psfig{figure=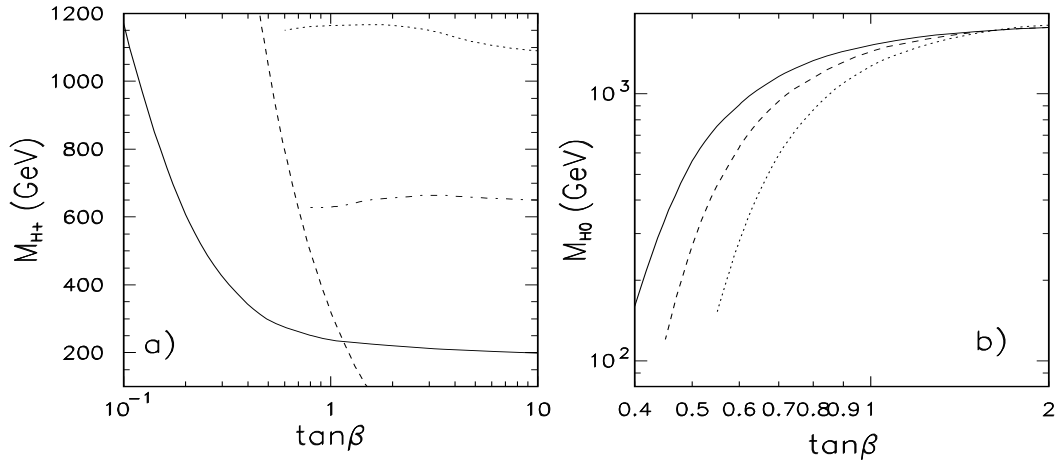,width=15.0cm,height=6.70cm} \vspace{1.0truecm}
\caption{{\bf a)} Lower limits on $H^+$ mass coming separately from 
$b\rightarrow s\gamma$ (solid line) and from  the requirement that 
$R_b^{2HDM}>R_b^{EXP} - 2(\Delta R_b)^{EXP}$  (dashed line)    as  a  
function   of
$\tan\beta$.  Also  shown are  upper limits  on $M_{H^+}$ arising  for
$M_h<20$ GeV from the requirement  of fine tuning in $M_{A}$  not
larger than     1\%  and   3\%    (dotted  and    dash-dotted   lines,
respectively). {\bf  b)} Upper limits on  $M_H$ in the case of light
$h^0$ ($M_h\leq$20--30 GeV; $\sin(\beta-\alpha)=0$) and   
$M_{H^+}=1000$,  800 and  
600    GeV  (solid, dashed and dotted lines, respectively).}
\label{fig:lhglobal}
\end{figure}

To  explain  the pattern   of $\chi^2$  seen in Figs.~\ref{fig:lhsmtb}a 
for $\tan\beta=2$ (which is representative of intermediate values
of $\tan\beta$ for which the predicted value of $R_b$  is as in the SM), 
recall  that the SM-type contribution (\ref{eqn:drhosm}) to $\Delta\rho$,
Eq.~(\ref{eqn:drho}), is negative and (for $\sin^2(\beta-\alpha)=0$)
decreases  with increasing $H^0$ mass.  This too negative contribution
can be easily compensated for by the contribution of the other Higgs 
bosons to $\Delta\rho$, when the equality  of $H^0$, $H^+$ and $A^0$ masses
is relaxed. From Eqs.~(\ref{eqn:drho}) and (\ref{eqn:drhoc})  it   follows   
that their contribution  is   positive,  provided $M_{H^+}\simgt M_A$.
Obviously, the $M_{H^+}$-$M_A$ mass
splitting must increase with increasing $M_H$ in order
to   compensate  for the increasingly negative   $H^0$ contribution  to
$\Delta\rho_{SM}$.  It  is also clear that   the possibility to adjust
the total $\Delta\rho$ to a proper    value  should hold also    for
larger $M_h$ and/or $\sin^2(\beta-\alpha)\neq0$.  We have checked that, 
for example for $M_h=50$ GeV and $\sin^2(\beta-\alpha)=0.2$ (see 
Fig.~\ref{fig:sinlim}), the plots look very similar to those for $M_h=20$ 
GeV and $\sin^2(\beta-\alpha)=0$. 

Since the contribution of $H^+$, $A^0$ and $h^0$ to $\Delta\rho$ depends  
on differences  of the masses  squared (quadratic violation of the $SU_V(2)$ 
``custodial''  symmetry), the cancellation between their contribution and 
$\Delta\rho_{SM}$ which is needed in
the case of $M_h\simlt20$ GeV to adjust $\Delta\rho$ to a proper value,
becomes more and more delicate as the mass of $M_{H^+}$ increases. This
leads to a stronger and  stronger correlation of $M_{H^+}$  with $M_A$,
clearly  seen   in Figs.~\ref{fig:lhsmtb} (note the different mass scales 
in the panels).  This  correlation   becomes particularly
fine-tuned in the case  of small $\tan\beta$,  where the requirement of
good $R_b$ forces $M_{H^+}$  to be large. The lower limits imposed on
$M_{H^+}$ by $b\rightarrow s\gamma$ and $R_b$ are shown  in 
Fig.~\ref{fig:lhglobal}a  by the solid and dashed lines, respectively. 
Since  the very  strong correlation  of $M_{H^+}$  with $M_A$ may seem
unnatural, it    is   interesting to see    how   the  requirement  of
``naturalness'' of the  $\chi^2$ constrains the parameter  space. This
is illustrated  by  the dotted   (dash-dotted) line in  
Fig.~\ref{fig:lhglobal}a,  which bounds from below
the region in  the $(\tan\beta, ~M_{H^+})$  plane in which a
change by $\simlt$1\% (3\%) of $M_A$ value which gives the minimum of
$\chi^2$ (for that particular point in the  $(\tan\beta, ~M_{H^+})$ plane), 
does  not   lead to
$\chi^2>19.5$. In producing these curves, the fit was always optimized 
with respect to the values of $M_H$ and $m_t$.
\footnote{Whenever we optimize with respect to $m_t$, the top mass 
measurement $m_t=(173.9\pm5)$ GeV \cite{TOPMASS} is included as  one of 
the fitted data in our $\chi^2$ fit.} 
It is  also interesting to note (see Figs.~\ref{fig:lhsmtb})
that in the case of heavy $H^+$, only
relatively  light $H^0$ can givea a good $\chi^2$  fit to the data. 
Therefore, for fixed values of $M_{H^+}$
and $\tan\beta$ the requirement    of $\chi^2<19.5$ leads to  an  {\it
upper bound} on $M_H$ shown in Fig.~\ref{fig:lhglobal}b (where we have
optimized the fit with respect to $M_A$ and $m_t$). 

\begin{figure}
\psfig{figure=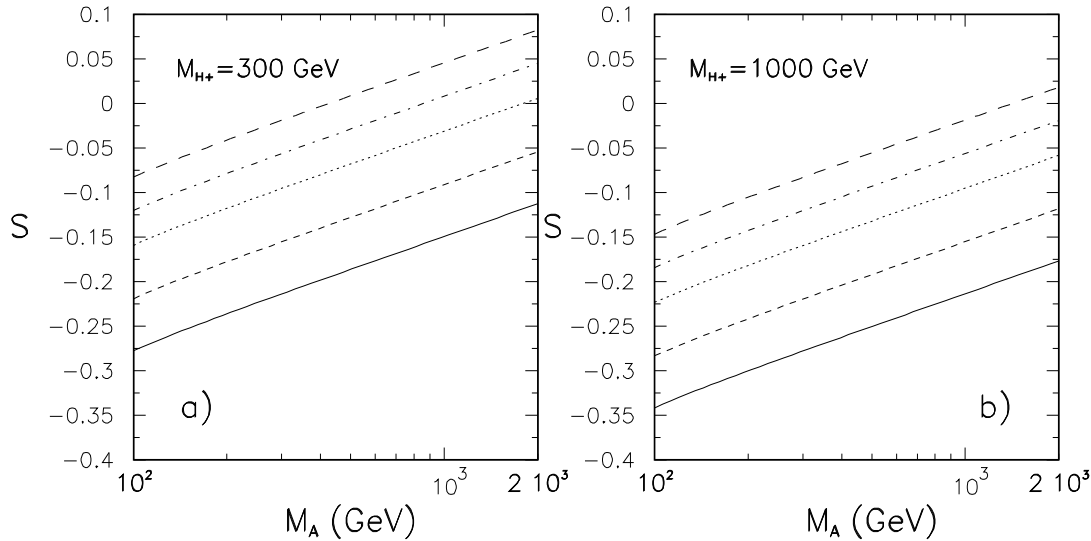,width=15.0cm,height=8.0cm} \vspace{1.0truecm}
\caption{Parameters $S$ as a function of $M_A$ for 
$M_{H^+}=300$ GeV and 1 TeV for $M_H=$ (from below) 100, 250, 500, 1000 and
2000 GeV.  In all cases $M_h=10$ GeV and $\sin^2(\beta-\alpha)=0$.}
\label{fig:lhparams}
\end{figure}

Larger value of  $\chi^2$ at the  minimum for heavier $H^0$ observed
in Figs.~\ref{fig:lhsmtb} can  be explained by the behaviour of the  
parameter $S$ \cite{STU}, which was not taken into account  in the 
above discussion. We define $S$ at   $q^2=M_Z^2$  rather  than  at 
$q^2=0$ (since it is $S(M_Z^2)$ that parametrizes more effectively   
the   electroweak observables measured at LEP):
\begin{eqnarray}
S = {4 s^2_W\over\alpha}\left[c^2_WF_{ZZ}(M_Z^2)-c^2_WF_{\gamma\gamma}
(M_Z^2) + {c_W\over s_W}(2s^2_W-1)F_{Z\gamma}(M_Z^2)\right],
\end{eqnarray}
where
\begin{eqnarray}
F_{ij}(q^2)\equiv{\Pi_{ij}(q^2) -\Pi_{ij}(0)\over q^2}.
\end{eqnarray} 
For $q^2=0$  it is easy to  obtain the analytic expression  for $S(0)$,
which we record in the Appendix.  Since for approximately fixed $M_A$
the parameter $S$ grows with increasing $M_H$, as shown in  
Fig.~\ref{fig:lhparams}, it is obvious  that for too
high values of $M_H$ the quality of the fit will be spoiled. 

When $M_{H^+}$ increases, its negative contribution decreases $S$,  
but this decrease is almost entirely compensated by the change in $M_A$,
which is required by the $\Delta\rho$ variable so that $\chi^2$ remains 
in its minimum (for that value of $M_{H^+}$). Therefore, the upper limit 
on $M_H$ is, for $\tan\beta\simgt1.5$, almost independent of the assumed value 
of $M_{H^+}$ as seen in Fig.~\ref{fig:lhglobal}b. For smaller $\tan\beta$, 
however, a lighter $H^+$ induces a larger negative contribution to 
$\delta R_b$, which 
has the effect that the upper limit on $M_H$ is stronger for
lighter $H^+$, since the minimum of $\chi^2$ is already higher than for 
heavier $H^+$. Therefore, this limit is rather stringent for 
$0.5\simlt\tan\beta\simlt0.8$. 
 
\begin{figure}
\psfig{figure=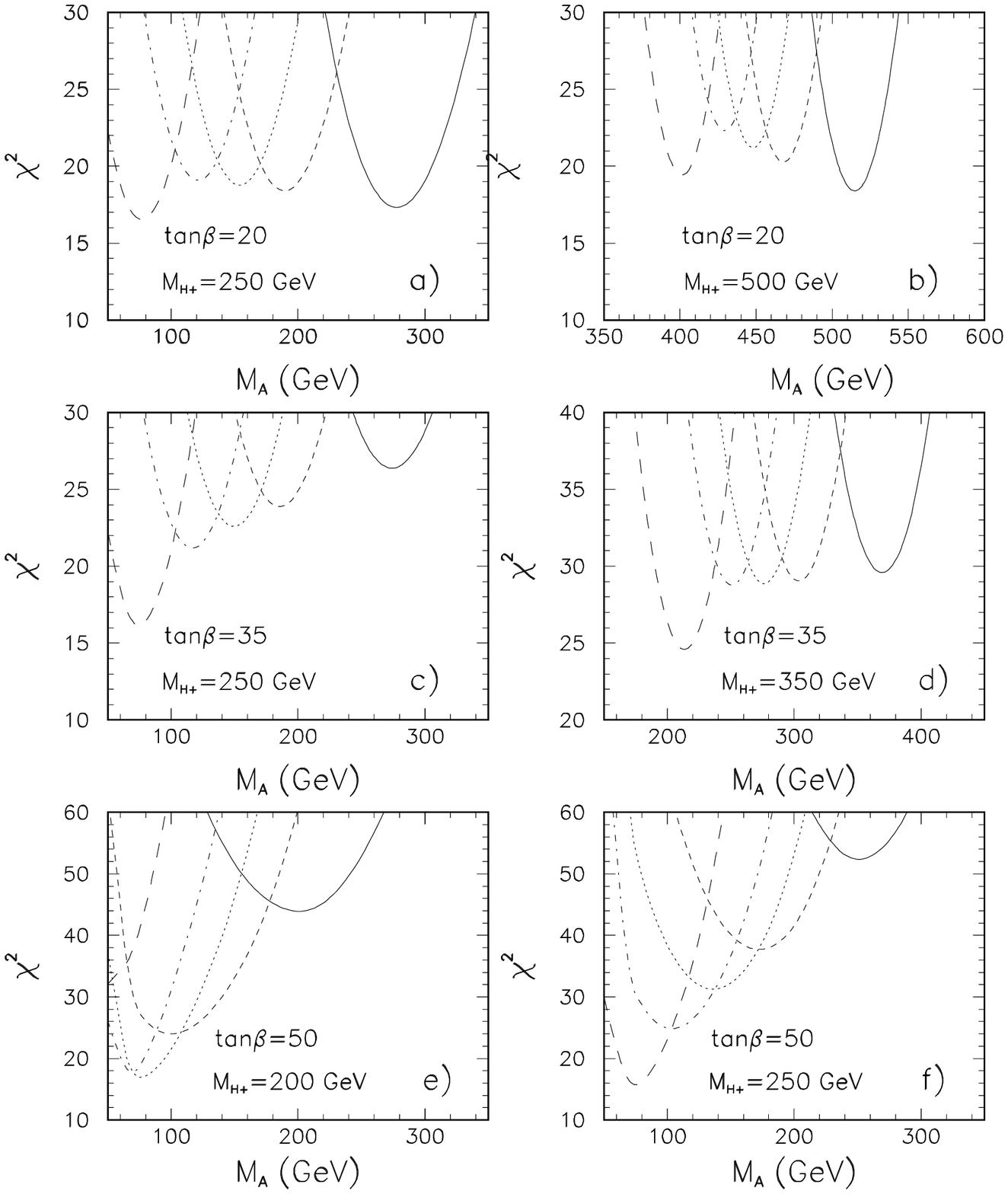,width=15.0cm,height=15.0cm}
\vspace{1.0truecm}
\caption{$\chi^2$ as a function of $M_A$ for three different 
values of $\tan\beta$ and different $H^+$ masses for $M_h=20$ GeV (i.e.
$\sin^2(\beta-\alpha)=0$). Solid, dashed, dotted, dot-dashed and 
long-dashed lines 
correspond to $M_H=$ 90, 500, 1000, 2000 and 5000 GeV, respectively.}
\label{fig:lhlargetb}
\end{figure}

\begin{figure}
\psfig{figure=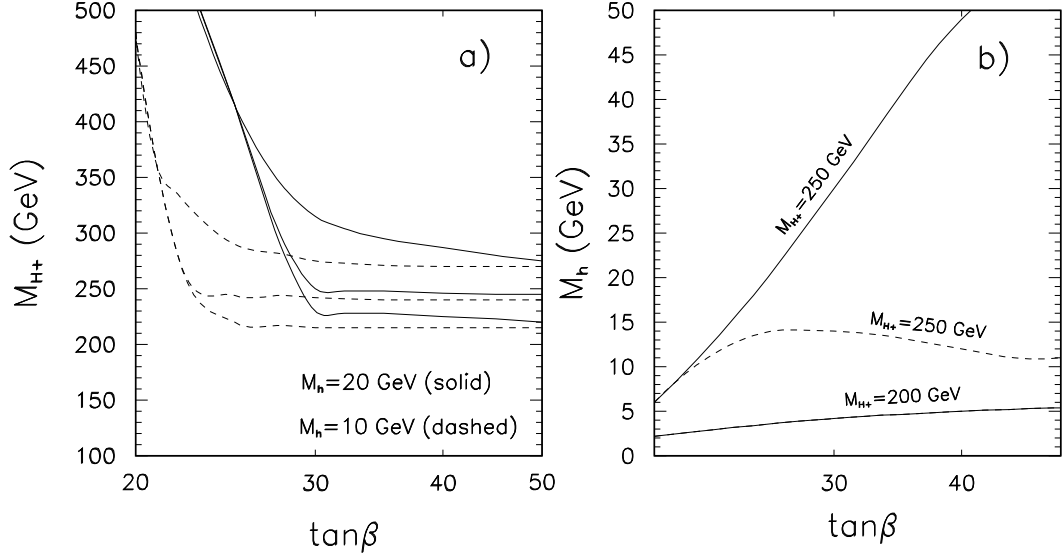,width=15.0cm,height=8.0cm}
\vspace{1.0truecm}
\caption{Limits from the $\chi^2$ fit: 
{\bf a)} Upper for the $H^+$ mass 
as a function of $\tan\beta$ for $M_h=20$ (solid lines) and 10 GeV 
(dashed lines) assuming the upper limit on
$M_H$ equal (from bottom to top) to 1, 3 and 5 TeV.
{\bf b)} Lower for $M_h$ for $M_{H^+}=$ 200 and 250 GeV (solid lines)
assuming the upper limit on $M_H$ equal to 1 TeV,
and for $M_{H^+}=$ 250 GeV (dashed line) and $M_H<3$ TeV.}
\label{fig:lhgloballtb}
\end{figure}

In  the case of  light $h^0$ and $\tan\beta\simgt20$, a qualitatively new
behaviour of $\chi^2$ appears as  a result of  the interplay of $h^0$,
$A^0$ and $H^+$ contributions to $R_b$ (for $\sin^2(\beta-\alpha)\approx0$ 
the $H^0$ boson contributes negligibly to  $R_b$). Because of the tight
correlation   of  $M_A$  with  $M_{H^+}$ (required    by $\Delta\rho$),
no acceptable $\chi^2$ can be   obtained for $H^+$  too heavy  since
there is then a large negative contribution to $R_b$ due to the large 
mass splitting between $h^0$ and $A^0$. Thus,  for large $\tan\beta$, very
light $h^0$ necessarily implies the existence of relatively light $H^+$. The
corresponding  pattern  of   $\chi^2$  (for fixed  $m_t=174$   GeV) is
illustrated in Fig.~\ref{fig:lhlargetb} for two different values of  
$M_{H^+}$ and $\tan\beta=20$, 35 and 50. Deeper minima of $\chi^2$ for 
larger values of $M_H$ seen in  Figs.~\ref{fig:lhlargetb}c--f can be 
explained by the  fact that (for fixed $M_{H^+}$)  larger  $M_H$ 
requires  lighter $A^0$ to   give acceptable  $\Delta\rho$ (the correlation 
seen already in Fig.~\ref{fig:lhsmtb}) and  this  increases the positive 
contribution to $R_b$ of the latter. {}From the pattern seen in 
Fig.~\ref{fig:lhlargetb}, it follows that there  exists, for a 
given mass of the lighter scalar $h^0$ and for a  given upper bound on
$M_H$, an upper bound on  $M_{H^+}$.  For $M_h=20$ and 10
GeV, this  bound (obtained for $\sin(\beta-\alpha)=0$  
by  scanning over $M_H$,  $M_A$ and $m_t$  and looking  for points with
$\chi^2<19.5$) is plotted in Fig.~\ref{fig:lhgloballtb}a, 
for an assumed upper limit on $M_H$ equal to
1000, 3000 and 5000 GeV. Taking  $h^0$ lighter strengthens the bound, as 
can be  easily inferred from Fig.~\ref{fig:tripledeg}. Of course,
for large $\tan\beta$  values allowing for  heavier $H^0$  weakens the
upper bound on $M_{H^+}$. However, for $\tan\beta$ values $\sim$20--30,
where $R_b$ gradually ceases to be so important (even allowing $M_H$ 
to vary up to 3 TeV), the minimum of
$\chi^2$  (for  given $M_{H^+}$) is  achieved for  $M_H=90$ GeV (lower
limit on the SM-like $M_H$ from direct searches). Therefore the minimum
of $\chi^2$ does not depend on the assumed upper limit on $M_H$. The 
importance of this upper bound on $M_H$ taken together with the lower 
one coming from $b\rightarrow s\gamma$  is obvious
\footnote{However, one should note that, for  given values of the Higgs
sector parameters, the minimum of $\chi^2$ for  $\tan\beta$ close to 50
occurs for $m_t\approx166$ GeV,
for which the  lower bound on $M_{H^+}$ coming from
$b\rightarrow  s\gamma$ is slightly  (by about  20 GeV) lower than the
bound plotted in Fig.~\ref{fig:lhglobal}a 
(which was obtained for $m_t=174$ GeV)}. 

Since the best $\chi^2$ is obtained, for $\tan\beta\simgt25$,  for the lowest
possible   mass of the   charged Higgs  boson and   $M_H$ equal to its
assumed upper  bound,  for a given value  of $\tan\beta$  there exists a
lower bound on  $M_h$ that follows  from  the assumed upper bound  on
$M_H$   and the lower  limit  on  $M_{H^+}$  coming from $b\rightarrow
s\gamma$. This  bound (obtained by scanning over $m_t$ and $M_A$ as well 
as over $\sin^2(\beta-\alpha)$ in the experimentally allowed range shown 
in Fig.~\ref{fig:sinlim}) is  shown  in Fig.~\ref{fig:lhgloballtb}b  for  
two different  lower bounds on $M_{H^+}$ taken to be 200 GeV (our limit
{}from $b\rightarrow s\gamma$) and 250 GeV and different upper
bounds on $M_H$. (For $M_{H^+}=200$ GeV, the upper limit on $M_h$ does not
decrease when larger $M_H$ are allowed, since it would require $M_A<65$ GeV,
which is excluded by the OPAL analysis, see Fig.~\ref{fig:mhmalim}.)
The bound turns out to be very sensitive to the lower
limit on $M_{H^+}$, showing that the  $b\rightarrow s\gamma$ process is
crucial for  constraining a light  scalar   Higgs scenario  for   large
$\tan\beta$. 

The global limits shown in Figs.~\ref{fig:lhgloballtb}a,b must be 
confronted with 
the recent analysis \cite{ZALEWSKI} of the Yukawa process for $h^0$ 
production. In the case of Fig.~\ref{fig:lhgloballtb}a parts of the solid
(dashed) lines corresponding to $\tan\beta\simgt40$ (30) seem to be
excluded by the data for the Yukawa process. However, those parts 
of these lines that are not excluded  still provide interesting and 
complementary limits on the light $h^0$ scenario.
In the case of Fig.~\ref{fig:lhgloballtb}b the limit obtained for $M_{H^+}=200$
GeV is for most of the $\tan\beta$ range weaker than the limit imposed by
the Yukawa process. For heavier $H^+$ these limits shown in 
Fig.~\ref{fig:lhgloballtb}b become competitive to the ones derived in 
ref. \cite{ZALEWSKI}.

\subsection{Light $A^0$}

In the the  case of light $A^0$  the analysis is more involved because
there is effectively   one more variable: $\sin^2(\beta-\alpha)$. 
Therefore let us begin again with    the  case    of  triple    degeneracy,
$M_h=M_H=M_{H^+}$.           As   is         clear     from       Eq.
(\ref{eqn:drho})  the  contribution  of   the  Higgs
sector to $\Delta\rho$  is in this limit  independent of the value  of
$M_A$  and  of     $\sin^2(\beta-\alpha)$; it is   given  simply   by
$\Delta\rho_{SM}(M_h)$. Therefore,  for $M_h=M_H=M_{H^+}$ and moderate
values of  $\tan\beta$ ($1.5\simlt\tan\beta\simlt20$), the best value of
$\chi^2$  depends on the lower  limit  on $M_{H^+}$ from $b\rightarrow
s\gamma$.  For $\tan\beta\simgt1$ the 
condition $\chi^2<19.5$  is satisfied if $M_{H^+}<240$
GeV, which is  still allowed (see Fig.~\ref{fig:lhglobal}). For
$\tan\beta\simlt1$ the light $H^+$ that is needed to give good $\Delta\rho$
tends to give too negative a contribution to $R_b$ 
(see Fig.~\ref{fig:lhglobal}a) and the condition $\chi^2<19.5$ cannot 
be satisfied.

\begin{figure}
\psfig{figure=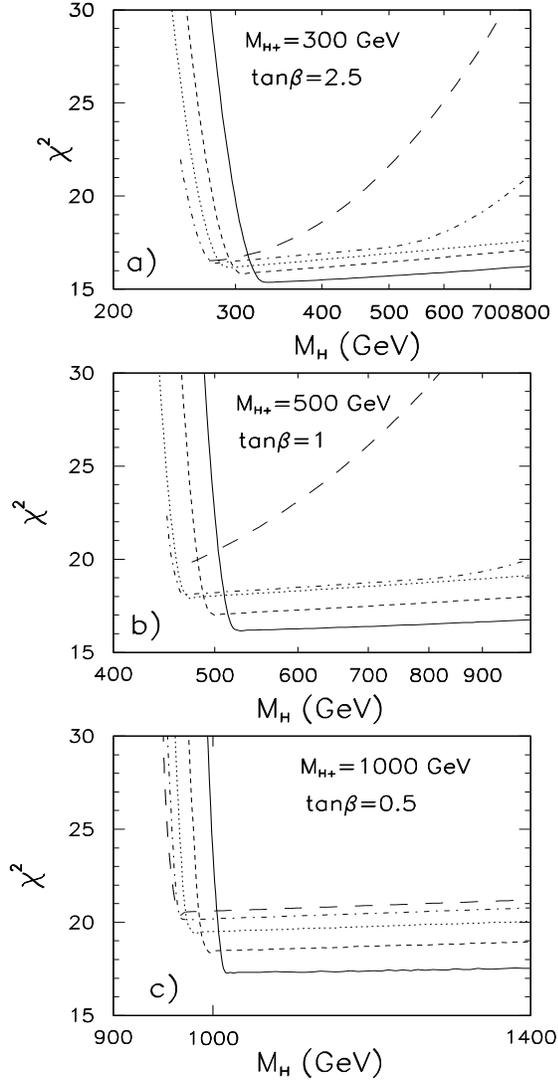,width=8.0cm,height=15.0cm} \vspace{1.0truecm}
\caption{$\chi^2$ as a function of $M_H$ for different low and intermediate
values of $\tan\beta$ and different $H^+$  masses, for $M_A=10$ GeV.
Solid, dashed, dotted, dot-dashed and long-dashed
lines correspond to $M_h$ equal (90, 150, 200, 250, 275) GeV, (90,
200, 400, 450, 475)  GeV and (90,  250, 500, 750,  940) GeV for panels
a, b and c, respectively.}
\label{fig:la}
\end{figure}

Even for $M_{H^+}$ larger
than 240  GeV, the value of $\chi^2$  can be kept below 19.5 by
the  departure from the   limit  of  triple degeneracy.  For  example,
consider the limit in which $h^0$ and $H^0$ remain degenerate (which makes
$\Delta\rho$ independent  of $\sin^2(\beta-\alpha)$). It is then easy
to see that for any value of $M_{H^+}$ (and any  $M_A$) there exists a
solution   to  the   equation     $\Delta\rho=0$  that occurs    for
$M_h=M_H\simlt  M_{H^+}$. 
For  values of $\tan\beta\simlt20$ (for  which neutral Higgs bosons
do not  play any role  in $R_b$) and $M_{H^+}$ not too large (so 
that the effects of the $S$ parameter are not too large, see below)
the existence of such    a  solution is sufficient to
ensure small values of $\chi^2$.  Note, however, that since the solution
to  the  equation $\Delta\rho=0$  is  due  to  the cancellation of  the
$\Delta\rho_{NEW}$, which depends on  the mass  splittings quadratically
against   $\Delta\rho_{SM}(M_h)$   which   depends  on    $M_h$   only
logarithmically, such a solution is strongly  fine-tuned (the more, the
heavier $H^+$).   The   fine-tuning  can be, however, reduced 
(or, more precisely, shifted to the variable $\sin^2(\beta-\alpha)$) by 
relaxing the  condition  $M_h=M_H$. This is illustrated in
Fig.~\ref{fig:la} where,  for  $M_A=10$ GeV   and  $m_t=174$ GeV, we show 
$\chi^2$  as a function  of  $M_H$  for  different (moderate and  low)
values of $\tan\beta$ and different choices of $M_{H^+}$ and $M_h$. In
all  cases   we optimize   $\chi^2$ with  respect    to the  value  of
$\sin^2(\beta-\alpha)$. 

The peculiar dependence  of $\chi^2$ as  a  function of $M_H$ seen  in
Fig.~\ref{fig:la} can be understood by inspection of the formulae for
$\Delta\rho$, Eq. (\ref{eqn:drho}). Consider first the case of $M_A=10$ 
GeV, $M_{H^+}=300$ GeV, as in Fig.~\ref{fig:la}a. For $M_h=90$  GeV  and 
$M_h\simlt M_H \ll M_{H^+}$, both factors,   $\Delta_c$ and
$\Delta_s$, are positive and obviously cannot cancel the positive first
term in Eq. (\ref{eqn:drho}) for   any  choice    of
$\sin^2(\beta-\alpha)$.      As $M_H$  increases, however,  $\Delta_s$
decreases rather  fast and  takes on   negative values for  $M_H\simlt
M_{H^+}$, whilst $\Delta_c$  decreases  slowly (logarithmically)  and
reaches  negative    values   only   for very large values of $M_H$.
Obviously,  for $M_H$  such that $\Delta_s<-\alpha A(M_A,M_{H^+})/4\pi
s^2_WM^2_W<\Delta_c$ 
there always exists   a  choice  of  $\sin^2(\beta-\alpha)$ for   which
$\Delta\rho = 0$.  This  explains  the plateau  in $\chi^2$ for  light
$h^0$ and  $M_H>M_{H^+}$. \footnote{Eventually, for $M_H$ large enough,
also $\Delta_c$ becomes smaller than $-\alpha A(M_A,M_{H^+})/4\pi s^2_WM^2_W$,
so that there is again no solution with $\Delta\rho=0$. This
happens for smaller $M_H$, the smaller the value of $M_{H^+}$.}
With $M_h$  increasing, $\Delta_c$ decreases
very  fast.   In    addition, $\Delta_s$ decreases   too,   though only
logarithmically.   As  a result, for  $M_h$  larger than some critical
value, both $\Delta_c$   and $\Delta_s$ become negative  and  smaller
than  $-\alpha A(M_A,M_{H^+})/4\pi s^2_WM^2_W$, and again $\Delta\rho$
cannot vanish  leading to large values of  $\chi^2$ seen in 
Figs.~\ref{fig:la}a-c for mass configurations corresponding to dot-dashed
and long-dashed lines. It should also be
obvious   that    for   very   heavy   $H^+$   the fine-tuning in
$\sin^2(\beta-\alpha)$   becomes extremely big,  making  such solutions
rather unnatural. Thus  light $A^0$ and values  of $\tan\beta\simlt1$,
for which $M_{H^+}$ must be large, are rather unlikely.

For very large  $M_{H^+}$, $\Delta\rho$ can vanish  only for heavy $H^0$. In
the  case of heavy $h^0$  this  means that the  parameter  $S$ is also
large (see Fig.~\ref{fig:lhparams}) and  spoils the  quality 
of the fit. This explains  why, in Fig.~\ref{fig:la}c, for 
larger values of $M_h$, and 
even for the other masses in  configurations for which $\Delta\rho$
can vanish by a judicious adjustment of $\sin^2(\beta-\alpha)$, the
value of $\chi^2$ is still above 19.5.

{}From the  above explanation   it   is clear    that for small    and
intermediate values  of $\tan\beta$ a  light $CP-$odd scalar $A^0$ can
be tolerated provided $h^0$ is lighter  than some bound $M_B$, which is
only slightly smaller than the mass of the  charged Higgs boson (which
in turn is constrained by $b\rightarrow s\gamma$ and, for $\tan\beta<1$, 
by the $R_b$ measurements).  At the  same time,  $M_H$ is bounded {}from
below also by roughly the same mass $M_B$. Of course, according to our
previous   discussion,  the case   $M_h=M_H\simlt  M_{H^+}$  is always
allowed (in this case $\sin^2(\beta-\alpha)$ is completely unconstrained).  
It is also worth noting  that, for $M_h\ll M_H\approx M_B\approx M_{H^+}$, 
the value of  $\sin^2(\beta-\alpha)$ (which is crucial for the $h^0$ and   
$A^0$ production processes) that is needed to keep $\chi^2$ below 19.5
is close to 1,  implying that $M_h$ is constrained,  in this case, 
by the LEP search to be greater than $\approx90$ GeV.
For heavier $H^0$, $\sin^2(\beta-\alpha)$
decreases, which means that the experimental lower bound on $M_h$ 
is also relaxed appropriately (recall that the limits from the associated
$h^0A^0$ production require only $M_h\simgt50$ (70) GeV for $M_A=10$ (50)
GeV - see Fig.~\ref{fig:mhmalim} - irrespectively of the value of 
$\sin^2(\beta-\alpha)$). The  decrease   of  $\sin^2(\beta-\alpha)$  with  
$M_H>M_{H^+}$  is slightly faster for heavier $A^0$  (and/or $h^0$) and 
slower for heavier $H^+$.

\begin{figure}
\psfig{figure=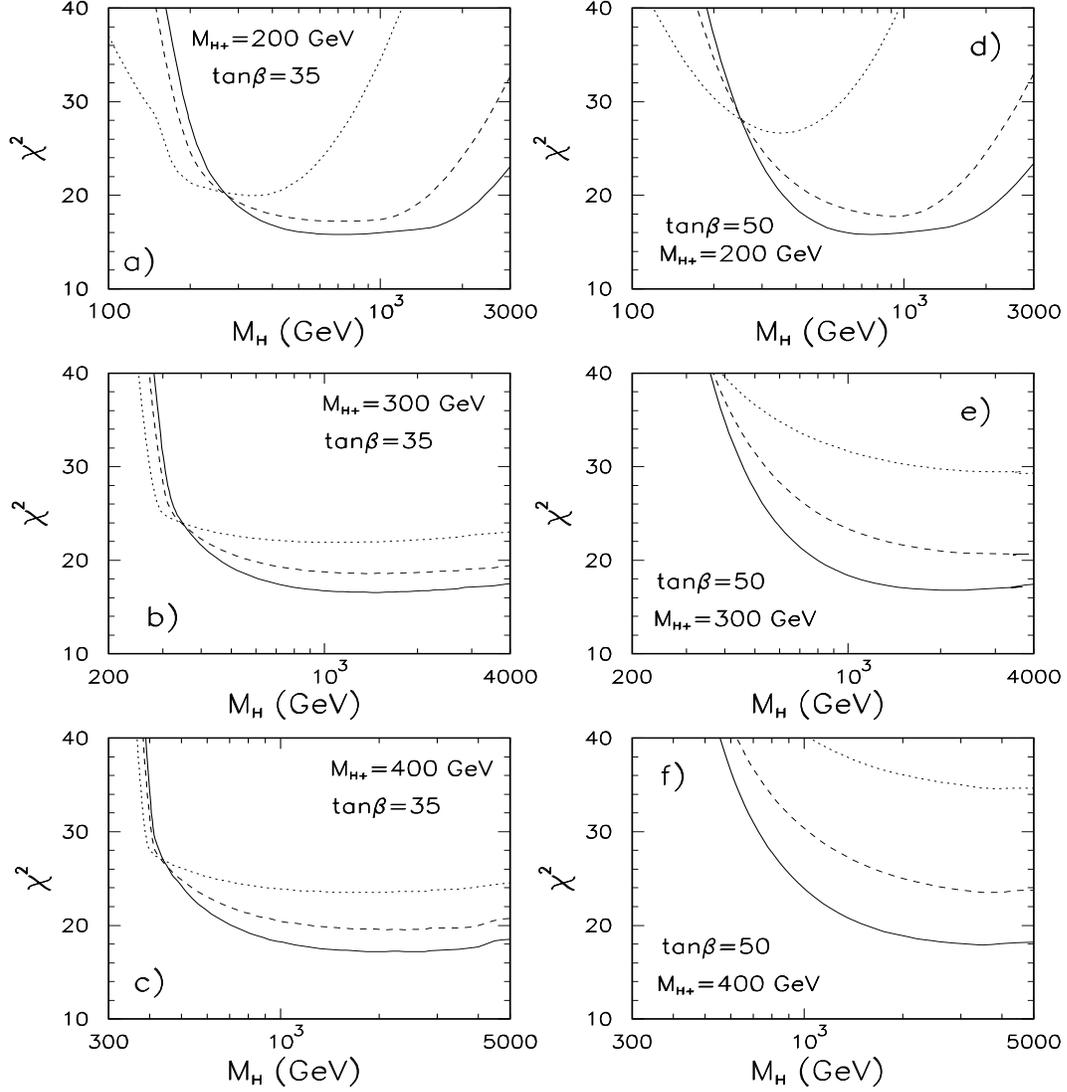,width=15.0cm,height=15.0cm} \vspace{1.0truecm}
\caption{$\chi^2$ as a function of $M_H$ for $\tan\beta=35$ and 50 
and different $H^+$ masses. Solid, dashed, dotted lines 
correspond to $M_h$ equal to 70, 90 and 150, respectively.
$M_A=25$ GeV.}
\label{fig:lalargetb}
\end{figure}

For large $\tan\beta$ the dependence of $\chi^2$ on $M_H$ for different 
values of $M_h$ and $M_{H^+}$ is shown in Fig.~\ref{fig:lalargetb}. The 
mass of the $CP-$odd scalar is taken to be 25 GeV so as to respect the 
bound from the Yukawa process also for $\tan\beta=50$ \cite{ALEPH,ZALEWSKI}. 
The behaviour of $\chi^2$ can be explained by
combining the information contained 
in Figs.~\ref{fig:la} (reflecting mainly the behaviour of $\Delta\rho$) with 
the behaviour of the corrections to $R_b$ for different Higgs boson mass 
configurations. In order for corrections to the latter quantity not to be 
too negative the neutral scalar that couples more strongly to the $b\bar b$ 
pair cannot be too heavy. On the other hand, for $M_H\simlt M_{H^+}$, a good 
$\Delta\rho$ is obtained for $\sin^2(\beta-\alpha)\approx1$ (i.e. 
$\sin\alpha\approx0$), which means that it is the mass splitting between $H^0$ 
and $A^0$ that is relevant to $R_b$. Only for sufficiently heavy $H^0$ does
$\sin^2(\beta-\alpha)$ become small enough ($\sin\alpha\approx1$) so that 
$h^0$ couples with full strength to $b\bar b$, and becomes relevant to $R_b$.
Thus, a good fit to the data can be obtained only with light $h^0$ and rather 
heavy $H^0$ (since this occurs for $\sin^2(\beta-\alpha)\ll1$, $h^0$ can be 
lighter than 90 GeV as we have just explained). The increase of $\chi^2$ for 
$M_H>1$ TeV seen in Fig.~\ref{fig:lalargetb}a and d for $M_h=70$ and 90 GeV 
is due to the fact that, for lighter $H^+$, $\Delta_c$ becomes smaller than 
$-\alpha A(M_{H^+},M_A)/4\pi s^2_WM^2_W$ already for relatively light $H^0$. 
Since, as explained above, $\Delta_c$ decreases very fast as $M_h$ increases, 
this effect is even more pronounced for $M_h=150$ GeV. Note that because of 
this effect for $M_{H^+}=200$ GeV one can reach $\chi^2<19.5$ only for 
$M_h<90$ GeV.

\begin{figure}
\psfig{figure=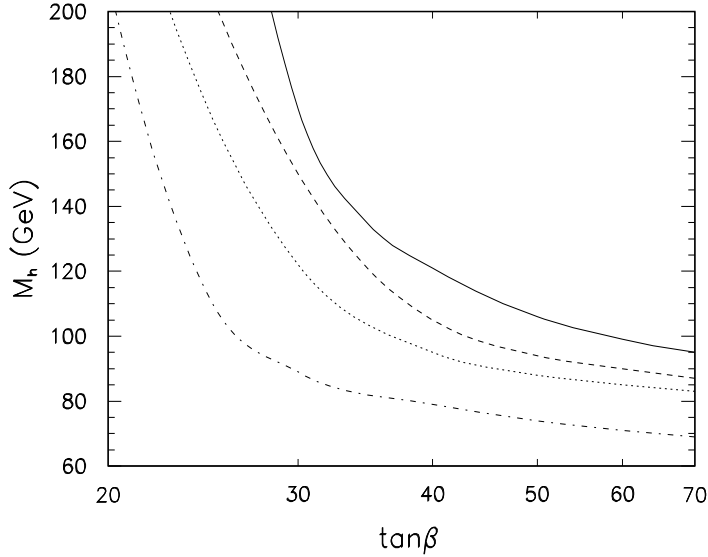,width=10.0cm,height=8.0cm} \vspace{1.0truecm}
\caption{Upper limit on $M_h$ in the light $CP-$odd scalar scenario
as a function of $\tan\beta$ for $M_A=$ 25 GeV (solid line), 15 GeV
(dashed line), 10 GeV (dotted line) and 1 GeV (dot-dashed line).}
\label{fig:alimh}
\end{figure}

For the same value of $M_H$ and $M_h$ the $\chi^2$ is slightly larger for 
heavier $H^+$ (despite the fact that the negative contribution of the latter 
Higgs boson to $R_b$ is decreased), because of the behaviour of 
$\sin^2(\beta-\alpha)$: it is larger for heavier $H^+$ and therefore, light 
$h^0$ does not fully compensate for the effects of light $A^0$ as $h^0$ 
couples more weakly to the $b\bar b$ pair. The conclusion following from the 
above considerations and from Figs.~\ref{fig:lalargetb} is that, in the 
scenario with large $\tan\beta$ and light $CP-$odd neutral Higgs particle
($M_A\simgt25$ GeV), the mass of the lighter neutral $CP-$even boson $h^0$
is bounded from above. This upper bounds on $M_h$ for four different 
values of the light $CP-$odd scalar mass (obtained by scanning over $m_t$,
$\sin(\beta-\alpha)$, $M_H$ and $M_{H^+}$) are shown as functions of 
$\tan\beta$ in Fig.~\ref{fig:alimh}. Of course, as $\tan\beta$ decreases,
the upper limit on $M_h$ approaches the bound $M_B\simlt M_{H^+}$ discussed
previously. The existence of an upper bound on $M_h$ means that 
LEP2 will be able to effectively test the scenario with light $A^0$
and large $\tan\beta$ either via the Bjorken process or the associated Higgs 
boson production, at least for $\tan\beta > 45 (30)$ for $M_A<25$ (10) GeV.

\vskip 0.3cm

\section{Summary of the results of the global fit and other constraints}

We have investigated the impact of the precision electroweak data 
on the parameter space of the 2HDM of type II. We have been particularly 
interested in constraints imposed on the very light neutral scalar 
scenarios (with $h^0$ or $A^0$ in the range $\simlt25$--30 GeV) by the 
requirement of good (as good as in the SM) fit to the electroweak data.
It turns out that neither scenario is excluded (directly or 
indirectly), provided some constraints are respected.

Apart from the well-known constraints on the charged Higgs boson mass
set in 2HDM(II) by $b\rightarrow s\gamma$ and, for $\tan\beta<1$, 
by $R_b$, we find that in the case of a light $CP-$even scalar $h^0$ 
and of intermediate values of $\tan\beta$ masses of the $CP-$odd and
charged Higgs bosons must be tightly correlated (the more, the heavier is
$H^+$) in order to maintain in the 2HDM(II) the same quality 
of the $\chi^2$ fit to the data as in the SM; this leads to strong 
``fine-tuning'' of this scenario. Therefore, 
limiting the acceptable degree of ``fine-tuning'' yields an upper bound on 
$M_{H^+}$ (Fig.~\ref{fig:lhglobal}a).
For a given mass of $H^+$ the requirement of a good fit to the 
data puts also an upper limit on $M_H$, which is particularly strong for 
$\tan\beta<1$ (Fig.~\ref{fig:lhglobal}b). 
For large values of $\tan\beta$ ($\tan\beta\simgt20$--30) the interplay
between the corrections to $\Delta\rho$ and $R_b$ implies that if $h^0$ 
is light, $H^+$ must also be light. The upper limit (as a function of 
$\tan\beta$) depends on the assumed upper bound on $M_H$ but, for $M_H$ 
in the TeV range, it is very strong for $\tan\beta\simgt35$, and $M_{H^+}$
of the order of
300 GeV (Fig.~\ref{fig:lhgloballtb}a). It is also interesting that, 
for large $\tan\beta$ and $M_{H^+}>200$ GeV, the electroweak data set the 
lower limit on the mass of $h^0$ shown in Fig.~\ref{fig:lhgloballtb}b.

For the light $A^0$ scenario we find that, for low and intermediate values
of $\tan\beta$, the quality of the fit is maintained provided 
$M_h < M_B < M_H$, where $M_B\simlt M_{H^+}$. For $\tan\beta>30$--35 there
emerges an additional upper bound on the mass of the lighter  scalar $h^0$,
which enables the test of the light $A^0$--large $\tan\beta$ configuration
at LEP2.

Other potential sources of further constraints on the 2HDM(II) that we 
have not considered here are the 
muon $g-2$ measurement \cite{MKZO} and $Z\rightarrow h(A)\gamma$ decays
\cite{MKMAZO}. 

The upper limit on $BR(Z^0\rightarrow h^0(A^0)\gamma)$ decays set by LEP1  
constrain our model only for $h^0(A^0)$ masses below 20 GeV and 
for values of $\tan\beta$ either very low (below 0.2) or very large,
above 55--70 GeV (which we have not considered). For larger $h^0 ~(A^0)$ 
masses, $\tan\beta$ is pushed outside the range (\ref{eqn:tbrange}) in 
which perturbative calculations 
can be done reliably. In this case, the resulting constraints 
can eventually be given a meaning as constraining the effective
(on-shell) $Z^0h^0\gamma$ and $Z^0A^0\gamma$ couplings but the relation
of these couplings to the original parameters of the model (which we have
been using) cannot be calculated perturbatively.

The constraints following from the present measurement of the $g-2$ of the
muon are stronger than the ones following from the Yukawa process only 
for $h^0$($A^0$) masses below 1--2 GeV, in which case they exclude
values of $\tan\beta\simgt4$ (10) for $M_{h(A)}=0.1$ (1) GeV. 
The E821 experiment may (depending on the measured central value) 
improve these limits  (according to the analysis presented in 
\cite{MKZO}), so that they become stronger than the Yukawa process ones
up to $M_{h(A)}\sim10$ GeV and exclude values of $\tan\beta\simgt2$ (15)
for $M_{h(A)}=1$ (10) GeV.

\vskip 0.3cm
\section*{Appendix}
Here we give the expression for the parameter $S(0)$. It differs slightly 
from the formula presented in ref. \cite{INLIYA}:
\begin{eqnarray}
S = -{1\over6\pi}\log{M_{H^+}\over M_W}
  +{1\over\pi}\sin^2({\beta-\alpha})
  \left[A^\prime(M_A,M_H)+A^\prime(M_W,M_h)+M^2_ZB^\prime(M_Z,M_h)\right.
\nonumber\\
+{1\over\pi}\cos^2({\beta-\alpha})
  \left[A^\prime(M_A,M_h)+A^\prime(M_W,M_H)+M^2_ZB^\prime(M_Z,M_H)
\right].\nonumber
\end{eqnarray}
The functions $A^\prime$ and $B^\prime$ are given by:
\begin{eqnarray}
A^\prime(m_1,m_2)={1\over12}\left[-{11\over6} 
+ {m_1^2\over m^2_1-m_2^2}\log{m^2_1\over M^2_W}
+ {m_2^2\over m^2_2-m_1^2}\log{m^2_2\over M^2_W}\right.\nonumber\\
+\left. {m_1^4+m_2^4\over(m^2_1-m_2^2)^2}
-{m^2_1m^2_2(m^2_1+m_2^2)\over(m^2_1-m_2^2)^3}\log{m^2_1\over m^2_2}\right],
\nonumber
\end{eqnarray}
\begin{eqnarray}
B^\prime(m_1,m_2)= -{1\over2}{m_1^2+m_2^2\over(m^2_1-m_2^2)^2}
+ {m^2_1m^2_2\over(m^2_1-m_2^2)^3}\log{m^2_1\over m^2_2}.\nonumber
\end{eqnarray}

\vskip 0.5cm

\noindent {\bf Acknowledgments}
\vskip 0.3cm
\noindent P.H.Ch.  has been partly supported by the Polish 
State Committee for Scientific Research grant 2 P03B 030 14
(for 1998--1999).  M.K. has been partly supported by the 
U.S.-Polish Maria Sk\l odowska-Curie Joint Fund II (MEN/DOE-96-264)
and by the Polish State Committee for Scientific Research 
grants 2 P03B 014 14 and 2 P03B 184 10.

We would like to thank F. Borzumati, C. Greub and M. Misiak for discussions. 
M.K. would also like to thank J. Gunion and W. Hollik for important
suggestions, H. Haber for discussions and hospitality during her stay at 
the Santa Cruz University, and K. Desh for providing some experimental data.

\end{document}